# MEASURING ORGANIZATIONAL CONSCIOUSNESS THROUGH E-MAIL BASED SOCIAL NETWORK ANALYSIS

Peter Gloor, Andrea Fronzetti Colladon
MIT, University Rome Tor Vergata
Cambridge USA, Rome Italy
pgloor@mit.edu, fronzetti.colladon@dii.uniroma2.it

**ABSTRACT**

This paper describes first experiments measuring organizational consciousness by comparing six "honest signals" of interpersonal communication within organizations with organizational metrics of performance.

**INTRODUCTION**

Ever since French enlightenment philosopher Rene Descartes put human consciousness into the sentence "Cogito ergo sum" – "I think therefore I am," researchers have been grappling with what human consciousness really is. In this paper we extend individual consciousness to collective consciousness, trying to identify communication patterns that might be indicative of the consciousness of groups. Teilhard de Chardin introduced the concept of "noosphere", the sphere of human thought complementing the biosphere, as the notion of a globally connected intelligence. Recently the concept of the "noosphere" has been gaining some traction, for instance by the Global Awareness Project at Princeton, which is trying to measure it using a network of sensors spread around the globe. The Princeton researchers claim to have identified a correlation between recognizable signals measured by their instrument, and significant external events such as earthquakes, or when the two airliners hit the World Trade Center 9-11[1]. Other researchers claim to have discovered traces of global consciousness for example in the Twittersphere (Dodds et al. 2011). In this paper we define organizational consciousness as common understanding on the team's global context, which allows team members to implicitly coordinate their activities and behaviors through communication (Daassi & Favier, 2007).

**SIX HONEST SIGNALS OF COMMUNICATION**

Our work focuses on organizations, and measures "honest signals" of communication, aggregating group consciousness along three dimensions of social interaction (figure 1): network structure, where we measure the degree of connectivity, dynamic changes in the network structure over time, where we measure the degree of interactivity of an actor, and content, where we measure the degree of sharing in word usage, sentiment, and emotionality.

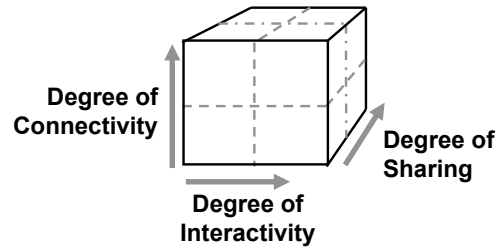

*Figure 1: Three-dimensional framework of group consciousness*

These three dimensions are the result of research conducted over the last twelve years analyzing hundreds of organizations on the global level, organizational, and individual level (www.ickn.org). On the global level we have studied dynamic networks constructed from re-tweets on twitter, link structure on Blogs, and co-authorship and Wikipedia link structure on Wikipedia. On the organizational level we have studied communication within dozens of organizations through their e-mail archive. On the individual level we have identified similar patterns using sociometric badges measuring interpersonal interaction.

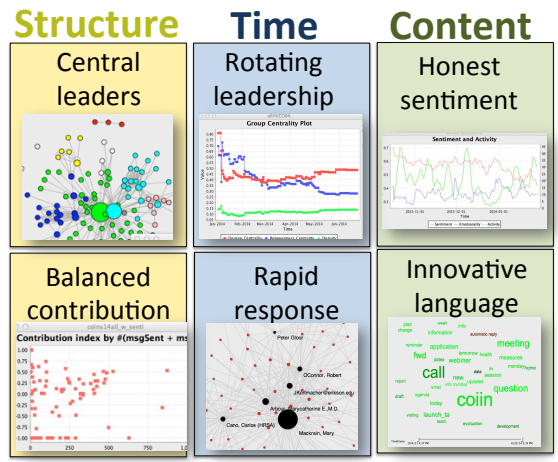

*Figure 2: Six honest signals of communication*

---

[1] http://noosphere.princeton.edu/results.html

On the structure, temporal, and content level we have identified six signals, which are excellent predictors of organizational creativity and performance (figure 2).

(1) Central leadership is measured through SNA metrics like group degree and group betweenness centrality.

(2) Balanced contribution is measured as the variance in contribution index, i.e. the ratio of sending to receiving messages.

(3) Rotating leadership is measured as oscillation in betweeness centrality and contribution index of actors.

(4) Rapid response is measured as the average time it takes an actor to respond to another, and the number of nudges it takes until the other actor responds.

(5) Honest sentiment is measured by the standard deviation in emotionality

(6) Innovative language is measured as the deviation in word usage from a standardized dictionary.

These six signals can be used to measure organizational consciousness.

## OPTIMIZING THE FLOW OF KNOWLEDGE

The six signals can then be calibrated with a dependent variable of organizational performance in a four-step process towards measuring organizational consciousness (figure 3).

In the first step, an e-mail-based structural social network analysis of the organization provides initial insights into key questions at the divisional, departmental, and role/individual level such as: Who are key influencers? Who is central in the network? How do they behave? Do they contribute to discussions or filter them? Do they assume a collegial/creative work style? Do they respond quickly? What is the sentiment of their conversations? How do business units interact with the rest of the organization? How do outside partners interact with the organization? Answering these ad similar questions assists in developing the hypotheses for the calibration step.

In the second step the six honest signals of communication are calculated.

In the third step, if the organization has performance metrics, these can be used evaluate the organization and ascertain whether certain communication patterns are associated with superior performance coefficients of the six honest signals, using or instance a regression model (see table 1 for an example).

In the fourth step, these organizational signals of consciousness can be continuously calculated, and interventions be taken to increase organizational effectiveness and creativity.

Table 1 illustrates analyzing organizational consciousness of a large company, where the six honest signals of communication of 16 independently operation business units are regressed against an exogenously measured organizational performance variable.

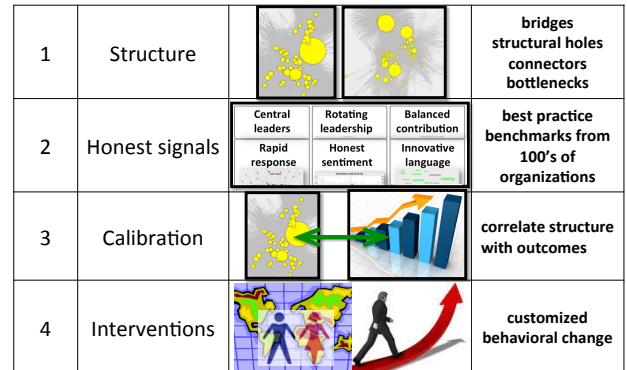

| 1 | Structure | bridges structural holes connectors bottlenecks |
| 2 | Honest signals | best practice benchmarks from 100's of organizations |
| 3 | Calibration | correlate structure with outcomes |
| 4 | Interventions | customized behavioral change |

*Figure 3: Four steps of knowledge flow optimization to measure organizational consciousness*

As the regression shows, the more emotional and responsive, and the less hierarchically structured a business unit is, the more successful it is.

|  | Model 1 Coeff. | Model 2 Coeff. | Model 3 Coeff. |
| --- | --- | --- | --- |
| **Predictors** | | | |
| Intercept | 0.1193777 | 0.6590365* | 1.064805** |
| Emotionality | 0.1409307* | 0.1409307** | 0.1228148** |
| Responsiveness |  | 0.0568062* | 0.0511518** |
| Structure |  |  | -0.0678522* |
| | | | |
| **FIT** | | | |
| N | 16 | 16 | 16 |
| Adj R2 | 0.2612 | 0.5163 | 0.6930 |

*Table 1. Regression results analyzing 16 organizational units of a company*

Through this approach we can substitute a dependent which is near impossible to measure – organizational consciousness – with measuring structural, temporal, and content based communication patterns, which, while still hard to measure, are far more tangible and measureable.